\documentclass[aps,prb,twocolumn,showpacs,amsmath,amssymb,superscriptaddress, citeautoscript]{revtex4-2}
\usepackage[utf8]{inputenc}
\usepackage[cmtip,all]{xy}
\usepackage{graphicx}
\usepackage{hyperref}
\usepackage{xcolor}
\usepackage{color}
\usepackage[all,cmtip]{xy}
\usepackage{multirow}
\usepackage{natbib}
\usepackage{amsmath,amsthm} 
\usepackage{amsfonts}

\usepackage{orcidlink}
\hypersetup{colorlinks=true, linkcolor=blue, citecolor=blue, urlcolor=blue, pdftitle={AM-topology}, pdfauthor={R. Gonzalez-Hernandez}}
\usepackage{xcolor}

\makeatother

\begin{document}
\par\noindent\rule{\textwidth}{0.5pt}
\title{Coexistence of d-Wave Altermagnetism and Topological States in Janus FeSeX (X = S, Te) Monolayers}

\author{Alvaro Gonz\'{a}lez-Garc\'{i}a\orcidlink{0000-0002-8789-0960}}\email{alvarogonzalez@uninorte.edu.co}
\affiliation{Departamento de F\'{i}sica y Geociencias, Universidad del Norte, Km. 5 V\'{i}a Antigua Puerto Colombia, Barranquilla 081007, Colombia}

\author{William  L\'{o}pez-P\'{e}rez\orcidlink{0000-0002-1460-4417}}
\affiliation{Departamento de F\'{i}sica y Geociencias, Universidad del Norte, Km. 5 V\'{i}a Antigua Puerto Colombia, Barranquilla 081007, Colombia}
	
\author{Paola Pacheco}
\affiliation{Grupo de Espectroscopia Optica de Emision y Laser, GEOEL, Universidad del Atlantico, Puerto Colombia, Barranquilla 081007, Colombia.}

\author{Luz Ram\'irez-Montes}
\affiliation{Departamento de F\'{i}sica, Universidad de Sucre, Barrio Puerta Roja, Sincelejo 700001, Colombia.}

\author{Rafael Gonz\'{a}lez-Hern\'{a}ndez\orcidlink{0000-0003-1761-4116}}
\affiliation{Departamento de F\'{i}sica y Geociencias, Universidad del Norte, Km. 5 V\'{i}a Antigua Puerto Colombia, Barranquilla 081007, Colombia}

\date{\today}

\begin{abstract}

The interplay between unconventional magnetism and band topology in two-dimensional materials has emerged as an important theme in condensed matter physics. Here, we present first-principles calculations that reveal the coexistence of $d$-wave altermagnetism and topological behavior in Janus FeSeX (X = S, Te) monolayers. The chemical asymmetry of the Janus structure breaks both out-of-plane mirror and inversion symmetries, leading to anisotropic exchange interactions and momentum-dependent spin splittings even in the absence of spin-orbit coupling, the defining signature of altermagnetism. 
Phonon dispersion analyses confirm the dynamical stability of both compounds, while strain-dependent calculations demonstrate that the magnitude of the altermagnetic exchange splitting ($\Delta_s$) can be efficiently tuned by biaxial strain. When spin-orbit coupling is included, a finite topological band gap emerges at the Fermi level, accompanied by quantized spin Hall conductivity plateaus and nontrivial topological invariants (spin Chern number = 1, $\mathbb{Z}_2 = 1$). These findings establish FeSeS and FeSeTe as promising two-dimensional platforms for realizing topological altermagnetism and spin--orbit--driven charge--spin conversion, thus opening new avenues for low-dissipation spintronic devices.
\end{abstract}

\maketitle	
\section{Introduction}

Altermagnetism has recently emerged as a distinct third class of magnetic order, different from both ferromagnetism and antiferromagnetism~\cite{smejkal2020sciadv,Smejkal2022,Smejkal2022a,mazin21,Gonzalez2021,Krempasky2024,bai2024altermagreview}. 
It features a magnetically compensated structure with zero net magnetization, but displays momentum-dependent spin splitting in the absence of spin–orbit coupling. 
This unique combination of magnetic compensation and spin polarization opens new avenues for dissipationless spin transport, spin–charge conversion, and the engineering of unconventional topological states~\cite{Jungwirth2025,ma2024altermagnetic,SCN_AM}. 
While most studies to date have focused on metallic altermagnets hosting nodal lines or Weyl points protected by magnetic  symmetries~\cite{MirrorChernBandsAM,nodalines-am,parshukov2025topological,AM-in-lieb-metal}, the realization of gapped topological altermagnets remains an active and rapidly developing area of research ~\cite{inducedmonolayeraltermagnetisfese,Jungwirth2026}.

In parallel, Janus two-dimensional (2D) structures, created by replacing one chalcogen or ligand layer in a layered compound, break out of plane mirror symmetry and enable functionalities such as piezoelectricity and tunable electronic structures that are absent in symmetric transition metal dichalcogenides (TMDs) \cite{gonzalez2024strong, zhang2017janus}. 
Fe-based TMD analogues have attracted significant attention due to their rich interplay between magnetism, superconductivity, and topology~\cite{chen2020synthesis,zhou2016interface,wang2014high,liu2012atomic,liu2012electronic,xiang2012high,Zhou2016,bafekry2021two}. 
Recent advances in the synthesis of ultrathin iron chalcogenides and Janus heterostructures have further broadened this research frontier, providing a versatile platform to explore symmetry-driven emergent phenomena in low-dimensional quantum materials~\cite{he2013phase,lai2015observation,paglione2010high}.

In particular, FeTe \cite{chen2020synthesis} and FeSe \cite{zhou2016interface} have been experimentally synthesized in the form of monolayers, nanoflakes, and transferable nanosheets. Although the hexagonal phase of Fe-chalcogenides is comparatively easier to synthesize than the tetragonal phase, recent experimental studies have successfully fabricated high-quality monolayer films of tetragonal Fe-chalcogenides \cite{qu2024high, he2013phase, lai2015observation, liu2012electronic}. For example, FeSe monolayer films obtained through an extensive annealing process have exhibited clear signatures of superconductivity with transition temperatures exceeding 65 K \cite{he2013phase}. Such achievements highlight the structural flexibility of Fe–chalcogen systems and establish a foundation for exploring emergent phenomena in reduced dimensionality. Furthermore, iron chalcogenides are especially compelling platforms for correlated and topological physics; notably, monolayer tetragonal FeSe exhibits interface-induced high-temperature superconductivity when grown in SrTiO${_3}$ \cite{song2019evidence, he2013phase, tan2013interface, liu2012electronic}.

Building on these experimental advances, density functional theory (DFT) studies have recently proposed Janus derivatives such as FeTeS and FeSeS \cite{Bafekry2021}. The asymmetric substitution of chalcogen in these compounds breaks inversion symmetry and is predicted to generate novel structural and electronic responses. Bafekry et al. \cite{Bafekry2021} explicitly noted that their predictions for Janus FeTeS and FeSeS were directly inspired by the experimental realization of FeTe${_2}$ monolayers \cite{chen2020synthesis}. Their calculations suggest that these Janus materials are dynamically stable, exhibit robust magnetism, and can even display a half-metallic behavior that is tunable by strain or electric field \cite{Bafekry2021}. However, their analysis overlooked the possible emergence of altermagnetic order and topological electronic phases.

Recently, 2D TMDs have been identified as fertile systems to induce altermagnetism in their intrinsic $P4mm$ symmetry \cite{khan2025altermagnetism, liu2025realizing, zhu2025two}. Khan et al. \cite{khan2025altermagnetism} demonstrated that this structural asymmetry in Janus TMDs produces spin-polarized electronic bands and anisotropic exchange interactions consistent with altermagnetic ordering. Liu et al. \cite{liu2025realizing} further revealed that strain and electric-field modulation can tune the magnitude and direction of symmetry-driven spin splitting, providing external control over altermagnetic behavior. Zhu et al. \cite{zhu2025two} extended this understanding by linking exchange interactions and lattice symmetry to the emergence of possible momentum-dependent spin textures. Regarding the feasibility of fabricating two-dimensional Janus FeSeX structures, the recent experimental growth of high-quality tetragonal FeSe flakes and monolayers on SrTiO${_3}$ substrates  \cite{song2019evidence, he2013phase, tan2013interface, liu2012electronic} demonstrates the possibility of manufacturing analogous two-dimensional Janus FeSeX structures. This progress suggests that such systems could be experimentally synthesized, providing a realistic platform to explore symmetry-protected altermagnetic and topological phases in two dimensions.

Although the field of 2D altermagnetic materials has grown rapidly \cite{khan2025altermagnetism, liu2025realizing, zhu2025two}, studies addressing gapped topological altermagnets remain relatively scarce. Here, we use first-principles theory to investigate the structural, magnetic, and topological properties of FeSeX (X=S, Te) Janus monolayers. Our findings  reveal that inversion-symmetry breaking, together with robust Fe exchange interactions, induces a compensated but momentum-dependent spin polarization, establishing an intrinsic $d$-wave altermagnetic phase. By including SOC, topological band inversion emerges, leading to quantized spin Hall effect responses and non-trivial topological invariants. Our results not only predict a new class of two-dimensional topological altermagnets but also provide a concrete materials platform in which altermagnetism and non-trivial band topology coexist. These results position Janus FeSeX monolayers as promising candidates for realizing symmetry-driven topological phases in two-dimensional magnetic materials.

\begin{table*}[!t]
	\centering
	\caption{Structural, magnetic, and electronic properties of Fe--chalcogenide monolayers and their Janus derivatives, obtained from density functional theory (DFT) and experimental data. Here SOC, M, SC, SM, STF denote Spin Orbit Coupling, Metallic, Semiconductor, Semimetal and  Semiconductor Thin Film, respectively.}
	\label{tab:ruo2_properties}
	\begin{tabular}{ccccc}
		\hline
		\textbf{ Material} & \textbf{$a$ (Å)} & \textbf{Method} & \textbf{Magnetic Properties} & \textbf{Electronic Properties} \\
		\hline
		\hline
         FeSe [This study] & 3.75 ($P4/nmm$) & DFT  & AFM & M \\
         FeSe ~\cite{liu2012electronic} & 3.77 ($P4/nmm$) & Epitaxial growth/SrTiO$_{3}$   & Superconductor & M \\
         FeSe ~\cite{long2020first} & 3.77 ($P4/nmm$) & DFT+U   & Nonmagnetic & M \\
         FeSe ~\cite{luo2022fragile} & 3.77 ($P4/nmm$) & DFT+U   &AFM  & Fragile topological band \\
         
        FeSeS [This study]   & 3.68 ($P4mm$) & DFT+U & AM & SOC-induced topological gap \\
        FeSeTe [This study]   & 3.82 ($P4mm$) & DFT+U & AM & SOC-driven topological insulator \\
        FeTe~\cite{Bafekry2021APL} & 3.81--3.86 ($P4/nmm$) & DFT / Exp. & FM/AFM & M \\
         FeSe~\cite{Zhou2016} & 3.67 ($P4/nmm$) & DFT & NM / weak AFM & SM; STF \\
FeS~\cite{Bafekry2021} & 3.68 ($P4/nmm$) & DFT & AFM & SC \\
        $h$-FeTe$_{2}$\cite{chen2020synthesis} & 3.695 ($P\bar{6}m2$)   & CVD & NM & M  \\
        h-FeTe$_2$~\cite{Bafekry2021} & 3.65--3.70 ($P\bar{6}m2$)  & DFT+U / Exp. & FM & M \\
       $h$-FeTe$_{2}$\cite{Bafekry2021} & 3.65 ($P\bar{6}m2$) & DFT+U & FM & M  \\       
		$h$-FeTeS \cite{Bafekry2021} & 3.41 ($P3m1$) & DFT+U  & FM & M\\ 
		$h$-FeSeS \cite{Bafekry2021} & 3.22 ($P3m1$)   & DFT+U   & FM & M\\ 
                
		\hline
		\hline
		\end{tabular}
        \label{Table 1}
\end{table*}

\section {COMPUTATIONAL DETAILS}
The first-principles calculations of the electronic structure were performed using the density functional theory (DFT) implemented in the \textsc{VASP} code ~\cite{Kresse1996}, using the PBE generalized gradient approximation ~\cite{Perdew1996}. Spin-polarized and noncollinear configurations were employed to describe the intrinsic antiferromagnetic ordering of the Fe sublattices. A plane-wave basis set with an energy cutoff of 460~eV and a Monkhorst--Pack mesh $\Gamma$-centered $15 \times 15 \times 1$ was used to ensure full convergence of the total energies and magnetic moments. Spin--orbit coupling (SOC) was included self-consistently. 

Electronic correlations within Fe--3$d$ orbitals were treated using the DFT+$U$ formalism in the Dudarev approach ~\cite{Dudarev1998}, with a moderate effective on-site interaction parameter comparable to that used by Mazin $et$ $al$. for monolayer FeSe ~\cite{inducedmonolayeraltermagnetisfese}, ensuring a consistent description of the  electronic and magnetic properties of Fe--3$d$ in Fe--based materials ~\cite{ekahana2024anomalous}.

The structural and magnetic symmetries of monolayer FeSe, FeSeS and FeSeTe were determined using the \textsc{Findsym} code \cite{stokes2005}. 
The electronic band structures were visualized with \texttt{PyProcar}\cite{pyprocar}, facilitating detailed orbital and spin-resolved analysis. 
To confirm the dynamical stability of both compounds, phonon dispersion calculations were performed using the \texttt{Phonopy} code\cite{phonopy} within a $3 \times 3 \times 1$ supercell.

Maximally localized Wannier functions were constructed from Fe--3$d$ and Se--$p$, Te--$p$, and S--$p$ orbitals using the \textsc{wannier90} package~\cite{Mostofi2014} to obtain low-energy tight-binding Hamiltonians. Based on these Wannier-interpolated models, the intrinsic spin Hall conductivity (SHC) was evaluated within the Kubo--Berry formalism on dense $160 \times 160 \times 1$ $k$-point grids. The energy-dependent SHC and the topological $\mathbb{Z}_2$ index were subsequently computed using the \textsc{wanniertools} code~\cite{Wu2018}, allowing a detailed characterization of the topological phase in the FeSeX Janus monolayers.

\section{RESULTS AND DISCUSSION}

\subsection{Fe-chalcogenide and Janus derivates}

The structural properties of monolayer FeSe and its Janus derivatives were investigated through first-principles DFT calculations. Pristine FeSe crystallizes in a tetragonal lattice with in-plane lattice constants \( a = b = 3.747~\text{\AA} \). The obtained lattice parameters are in excellent agreement with previous experimental \cite{liu2012electronic} and theoretical \cite{long2020first, luo2022fragile} reports, as summarized in Table~\ref{Table 1}, confirming the accuracy of our structural optimization.

In pristine FeSe, Fe atoms occupy the Wyckoff position \( 2b \) and exhibit a collinear antiferromagnetic configuration with out-of-plane magnetic moments of \( 2.2~\mu_B \), while Se atoms reside at the \( 2c \) sites, located above and below the Fe plane. This spin arrangement lowers the symmetry of the nonmagnetic space group \( P4/nmm \) (No.~129) to the magnetic space group \( P4'/n'm'm \) (No.~129.416).  The magnetic symmetry generators include the fourfold rotation combined with time reversal (\( C_{4z}\mathcal{T} \)), mirror reflections with time reversal and partial translations (\( M_x\mathcal{T}t_{x},\, M_y\mathcal{T}t_{y},\, M_z\mathcal{T}t_{x+y}\)),  and inversion (\( \mathcal{I} \)). 

The Janus \( \mathrm{FeSeX} \) (\( X = \mathrm{S}, \mathrm{Te} \)) monolayers are obtained by replacing one of the Se sublayers in pristine FeSe with S or Te atoms, introducing an intrinsic out-of-plane chemical asymmetry. 
This atomic modification breaks all operations that map $z$ to -$z$, including both the mirror reflection in the $z$-plane ($M_z\mathcal{T}t_{x+y}$) and inversion symmetries, while preserving the four-fold rotation symmetry. 
Consequently, the resulting magnetic Janus structures belong to the noncentrosymmetric magnetic space group \( P4'm'm \) (No.~99.165), characterized by the antiunitary four-fold rotation \( C_{4z}\mathcal{T} \) symmetry that connects the two alternating Fe magnetic sublattices. 
This symmetry is also preserved within the spin-space symmetry formalism and defines the altermagnetic nature of the Janus \( \mathrm{FeSeX} \) compounds.
The altermagnetic ground state was identified in both Janus FeSeS and FeSeTe by total-energy calculations (see Table S1 of the Supplemental Material \cite{SuppMat}).

Table~\ref{Table 1} summarizes the structural and magnetic properties of Fe–chalcogenide monolayers and their Janus counterparts, as obtained from DFT calculations and available experimental data. 
The results highlight the lattice distortions and symmetry breaking induced by Janus engineering, which differentiates the noncentrosymmetric \( P4mm \) phases of FeSeX from their centrosymmetric \( P4/nmm \) and hexagonal \( P\bar{6}m2 \) analogs. 
The lattice parameters of FeSe, FeSeS and FeSeTe exhibit slight variations associated with the electronegativity differences among S, Se, and Te. 
Pristine FeSe in the monolayer \( P4/nmm \) phase shows an in-plane lattice constant of \( 3.747~\text{\AA} \), while the Janus FeSeS and FeSeTe monolayers display \( 3.68~\text{\AA} \) and \( 3.82~\text{\AA} \), respectively. 
The progressive increase in lattice constants from S- to Te-containing FeSe layers is consistent with the decreasing electronegativity and increasing atomic radius of the chalcogen atoms

\subsection{Dynamical Stability} 
\begin{figure}[!t]
	\center
	\includegraphics[width=0.50\textwidth]{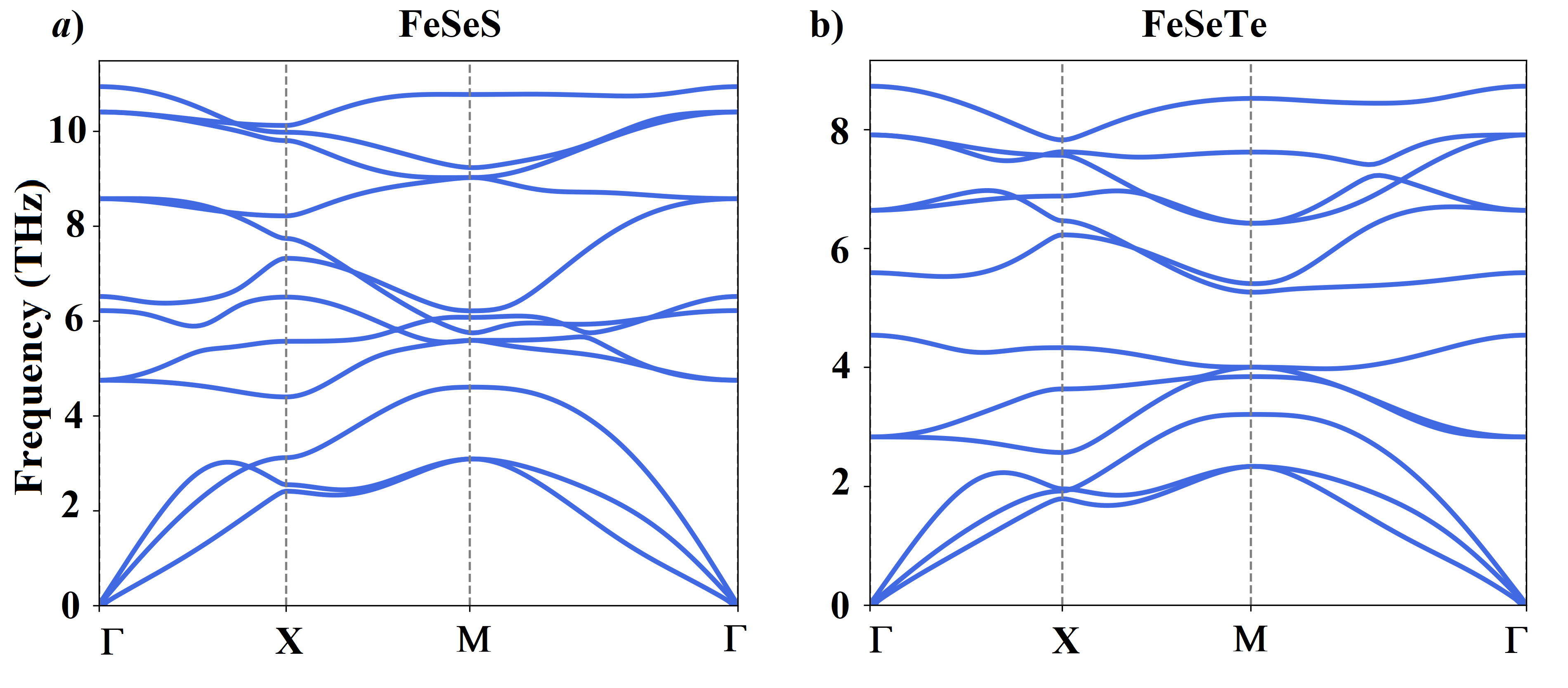}
	\caption{Phonon dispersion relations of a) FeSeS and b) FeSeTe along the high-symmetry path $\Gamma$--X--M--$\Gamma$ in the Brillouin zone. The phonon branches represent the vibrational modes of the Fe--chalcogen lattice, with acoustic modes emerging from $\Gamma$ and optical modes extending up to $\sim 11$ THz and $\sim 9$ THz for FeSeS and FeSeTe, respectively. The absence of imaginary frequencies throughout the Brillouin zone confirms the dynamical stability of both compounds.}
	\label{fig:ph}
\end{figure}

The phonon dispersion relations calculated along the high-symmetry directions of the tetragonal Brillouin zone, shown in Figure.~\ref{fig:ph}, confirm the dynamical stability of the Janus compounds \( \mathrm{FeSeX} \) (\( X = \mathrm{S}, \mathrm{Te} \)). 
Figure~\ref{fig:ph}(a) displays the phonon spectrum of FeSeS, which exhibits a broad frequency range with optical branches extending up to approximately $\sim 11$ THz, reflecting the stronger interatomic bonding and the lighter mass of sulfur compared to selenium. 
The separation between acoustic and optical branches indicates moderate phonon–phonon coupling and further supports the dynamic stability of the lattice, as evidenced by the absence of imaginary (negative) frequencies. 
The steep slope of the acoustic modes near the $\Gamma$ point suggests relatively high sound velocities, consistent with a mechanically robust and elastically stiff lattice.

In contrast, the FeSeTe compound shows a systematic redshift of the phonon frequencies, with optical branches confined below $\sim 9$ THz. This downward shift arises from the heavier Te atoms, which lower the vibrational frequencies due to the inverse square-root dependence on atomic mass. The flattening of several optical branches also indicates increased phonon scattering and reduced lattice stiffness. Despite these shifts, FeSeTe remains dynamically stable since all phonon modes are positive throughout the Brillouin zone. 

Beyond phonon analysis, the thermal stability of Janus 2D-FeSeX monolayers (X = S, Te) was further investigated by ab-initio molecular dynamics simulations (AIMD) performed within the NVT ensemble at room temperature (300 K). The results,  included in the Supplemental Material (Figure S1) \cite{SuppMat}, show that the temperature remains well stabilized around the target value, and the total energy shows only small fluctuations without noticeable drift during simulation. Furthermore, structural snapshots for both FeSeS and FeSeTe taken after the equilibration stage (0.5–5 ps) confirm that the structures remain stable at room temperature (Figure S2 \cite{SuppMat}). Snapshots show that the atomic configurations remain essentially unchanged throughout the simulation, with no evidence of bond breaking, structural reconstruction, or significant lattice distortion. The Fe–Se–X (X = S, Te) coordination environment is preserved throughout the AIMD trajectory, indicating that both monolayers maintain their structural integrity under thermal fluctuations at 300 K.


\begin{figure*}[!t] 
 	\centering
 	\includegraphics[width=0.860\textwidth]{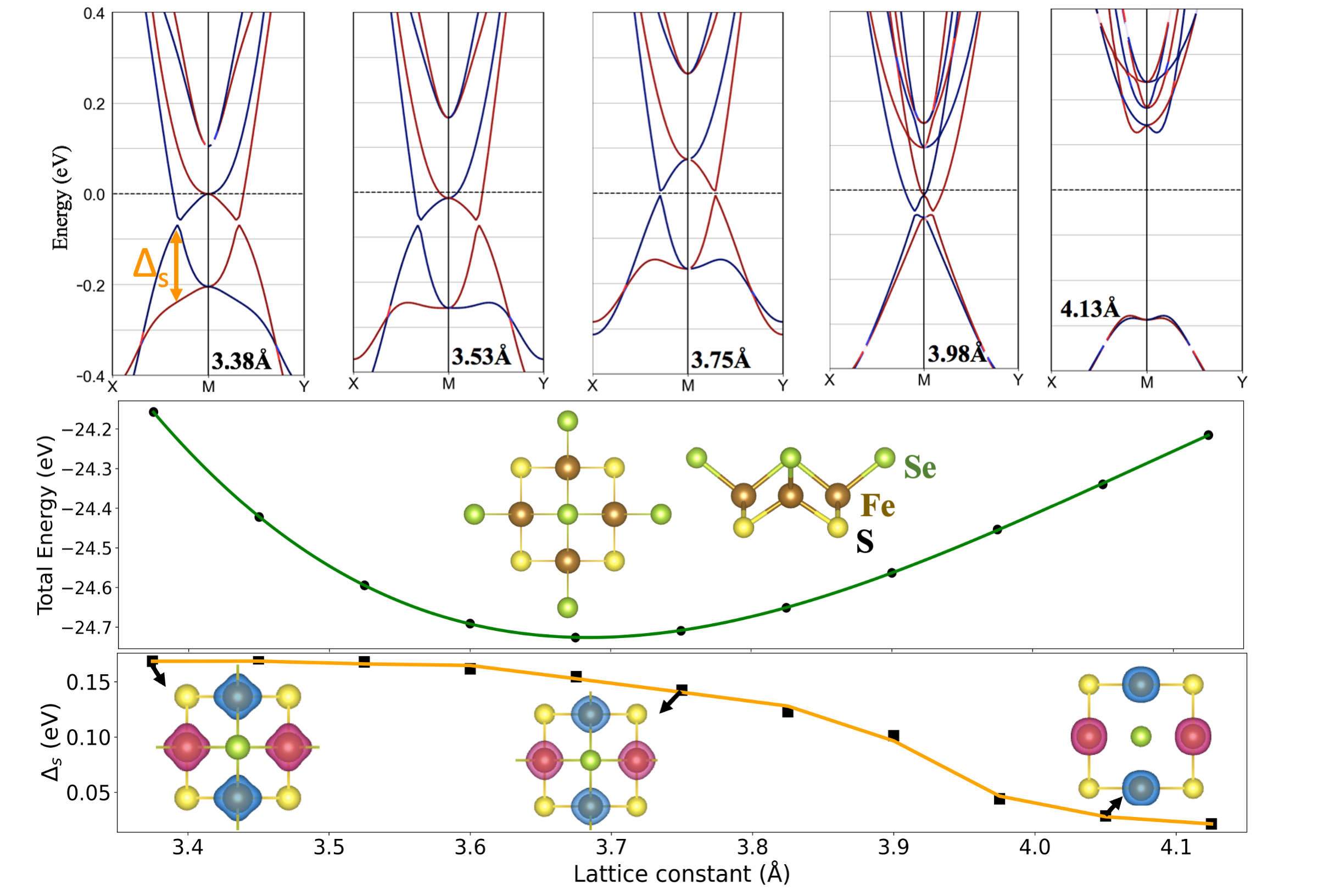}
 	\caption{(Color online) Strain-tunable altermagnetism in Janus FeSeS. (Top) Spin-polarized band structures for selected in-plane lattice constants \( a \), without SOC. Blue and red denote opposite spin channels; the orange arrow marks the exchange splitting \( \Delta_s \) near the M point. (Middle) Total energy vs \( a \) and structural schematic (Fe: brown, Se: green, S: yellow); the equilibrium \( a \) is $\sim$ 3.68 ~\text{\AA}. (Bottom left) Evolution of \( \Delta_s \)  with \( a \), showing enhancement under compression and suppression under tension. (Bottom right) Spin-density isosurfaces for three representatives \( a \) values, illustrating the emergence and collapse of the staggered spin density with strain.}
 	\label{fig:FeSeS}
 \end{figure*}

\subsection{Electronic structure}
In pristine FeSe monolayers, the electronic band structure remains spin-degenerate as a consequence of the antiunitary symmetry 
\(\mathcal{G}_z = M_z \, \mathcal{T} \, t_{x+y}\).
This operation combines a mirror reflection through the monolayer plane (\(M_z\)), time-reversal symmetry (\(\mathcal{T}\)), and a partial translation along the \(x+y\) direction (\(t_{x+y}\)).
The composite symmetry \(\mathcal{G}_z\) maps the two antiferromagnetically aligned Fe sublattices onto each other, ensuring that every electronic state has a degenerate  opposite spin throughout the 2D Brillouin zone (BZ).
This Kramers-like degeneracy can be expressed as
\begin{equation}
	\mathcal{G}_z \, E_{\uparrow}(k_x, k_y) = E_{\downarrow}(k_x, k_y),
\end{equation}
demonstrating that spin-up and spin-down bands are energetically equivalent at each $k$-point in the BZ.
Within the spin-group formalism, this symmetry is equivalent to [$C_2\parallel m_z$], which enforces spin degeneracy across the entire reciprocal space~\cite{2d-am-spingroup}. Figure~S6 illustrates this behavior, showing the FeSe band structures for different lattice constants, where spin degeneracy is preserved under both compressive and tensile strain (see  the Supplemental Material \cite{SuppMat}).

\begin{figure*}[!t]
 	\centering
 	\includegraphics[width=0.860\textwidth]{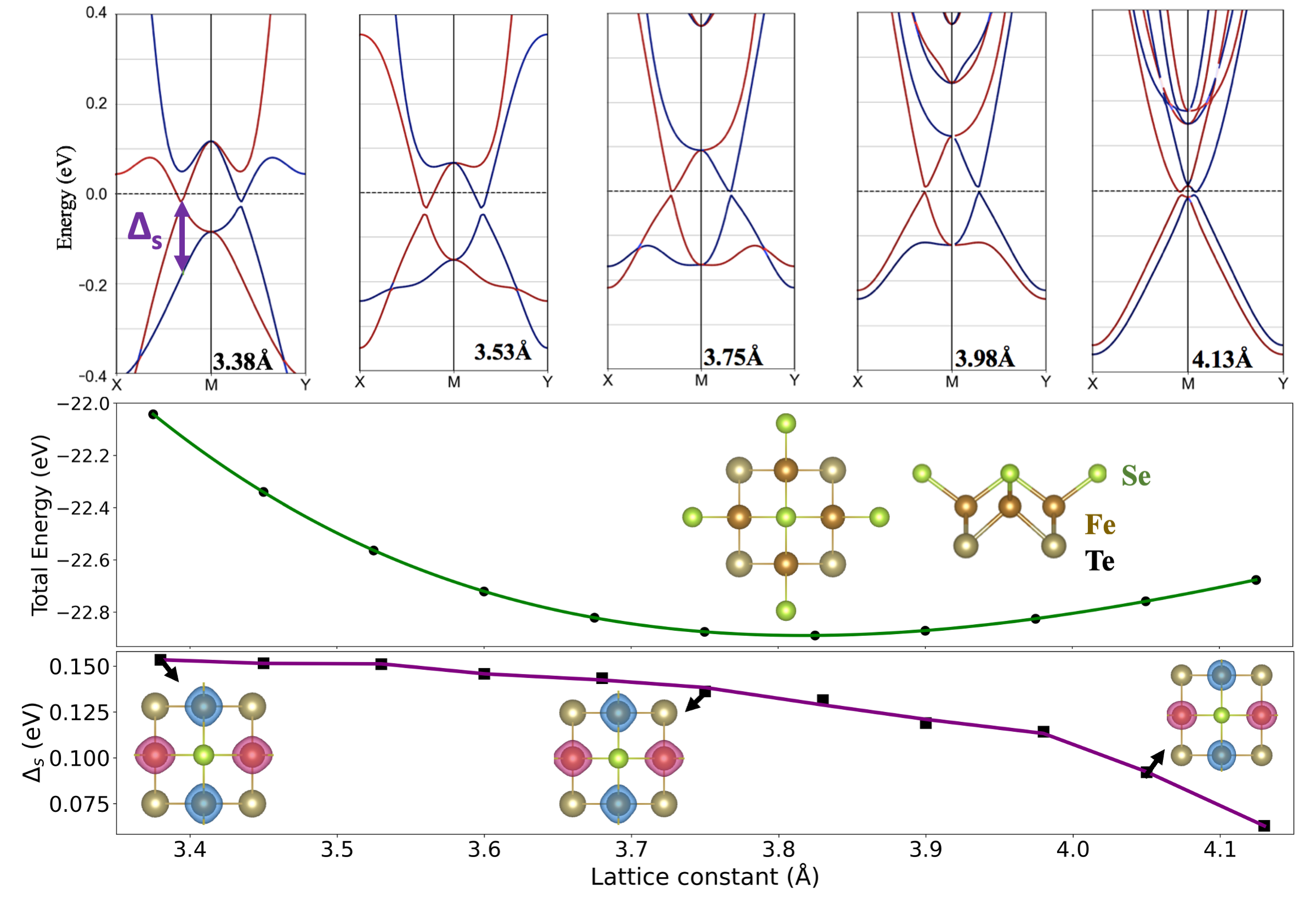}
 	\caption{(Color online) Strain-tunable altermagnetism in Janus FeSeTe. (Top) Spin-polarized band structures for representative in-plane lattice constants \( a \), without SOC. (Middle) Total energy vs \( a \). (Bottom) Evolution of the exchange splitting \( \Delta_s \)  near M and spin-density isosurfaces (spin-up blue, spin-down red) for compressed, equilibrium, and tensile states. Compressive strain enhances \( \Delta_s \) and stabilizes an staggered spin texture, while tensile strain suppresses the spin splitting and collapses the altermagnetic order.}   
 	\label{fig:FeSeTe}
 \end{figure*}

By contrast, Janus \( \mathrm{FeSeX} \) (\( X = \mathrm{S}, \mathrm{Te} \)) materials break the \( M_z\mathcal{T} t_{x+y}\) symmetry, removing the global Kramers degeneracy while retaining the antiunitary fourfold rotation \( C_{4z}\mathcal{T} \) to connect the opposite Fe sublattices.
This symmetry enforces band degeneracies along the \( C_{4z}\mathcal{T} \)-invariant lines (\( \Gamma \text{--} M \)) and induces altermagnetic behavior in the Janus \( \mathrm{FeSeX} \) systems, producing opposite spin splittings along \( M \text{--} X \) and \( M \text{--} Y \)~\cite{vsmejkal2022emerging}, as shown in the band structures of Figures~\ref{fig:FeSeS} and~\ref{fig:FeSeTe}  .

Figure~\ref{fig:FeSeS} shows the strain-dependent electronic band structure and energetic properties of the Janus FeSeS monolayer in the nonrelativistic limit. 
The top panels show the spin-resolved band dispersion along the high-symmetry directions for a representative in-plane lattice constant \(a\). 
For lower values of the equilibrium lattice constant (\( a \approx 3.68~\text{\AA} \)), the system exhibits metallic behavior with a finite, momentum-dependent spin splitting near the valence-band maximum, located around the \(M\) point in the BZ.
The maximum spin-splitting, denoted as \(\Delta_s\), occurs close to the \(M\) point and originates purely from the exchange interaction associated with the compensated magnetic configuration from the Janus geometry.

As shown in the bottom panel of Figure~\ref{fig:FeSeS}, the magnitude of \( \Delta_s \) exhibits a strain dependence. 
Under compressive strain (\( a < 3.68~\text{\AA} \)), enhanced Fe--chalcogen hybridization strengthens the exchange interaction, leading to a larger spin splitting of \( \Delta_s \approx 0.16\text{--}0.18~\text{eV} \) at \( a = 3.38~\text{\AA} \). 
Conversely, tensile strain reduces orbital overlap and suppresses the altermagnetic splitting to below \( 0.02~\text{eV} \) for \( a \approx 4.13~\text{\AA} \). 
This decrease demonstrates that strain serves as an effective control parameter for tuning the spin splitting in Janus FeSeS. 
Furthermore, the total-energy profile shown in the central panel reveals that the equilibrium lattice constant coincides with a sizable \( \Delta_s \), indicating a favorable balance between structural stability and the emergent $d$-wave altermagnetic order.

The bottom panel of Figure~\ref{fig:FeSeS} also displays the spin-density distribution for three representative lattice constants. 
The spin-density isosurfaces (spin-up in blue and spin-down in red) provide a real-space visualization of the microscopic origin of the momentum-dependent exchange splitting. 
At \( a = 3.38~\text{\AA} \), the spin density exhibits an anisotropic distribution localized on the two Fe sites, with opposite spin polarization. 
This real-space anisotropy gives rise to the $d$-wave altermagnetic order, in which the magnetic moments form a compensated configuration (zero net magnetization) but locally break symmetry, coupling spin density to crystal momentum~\cite{Smejkal2022b}. 
This broken local symmetry induces the observed spin splitting of \( \Delta_s \approx 0.17~\text{eV} \) in the band structure. 
At \( a \approx 3.68~\text{\AA} \), near the equilibrium lattice constant, the spin density retains its anisotropic character with a slight reduction in the effective exchange interaction and a smaller \( \Delta_s \), consistent with the trend in  band dispersion. 
Under large tensile strain (\( a = 4.13~\text{\AA} \)), the spin-density becomes nearly spherically symmetric, signaling the collapse of the altermagnetic state and the vanishing of \( \Delta_s \) in the electronic structure.  
In this regime, the system loses its metallic character and transitions to insulating behavior.

For the Janus FeSeTe monolayer, Figure~\ref{fig:FeSeTe} shows the strain-dependent evolution of its electronic and magnetic properties. 
The spin-resolved band structures (top panels), total energy as a function of the in-plane lattice constant \( a \) (middle), and the variation of the exchange-induced splitting \( \Delta_s \) together with representative spin-density isosurfaces (bottom) provide reciprocal- and real-space evidence that FeSeTe hosts a strain-tunable altermagnetic state. 
In the equilibrium geometry (\( a \approx 3.82~\text{\AA} \)), the spin density exhibits an anisotropic and staggered texture localized on the Fe atoms, giving rise to a momentum-dependent band splitting near the \( M \) point. 
This anisotropic, yet magnetically compensated, configuration constitutes the real-space signal of altermagnetism in 2D systems lacking mirror or inversion symmetry. 
Under compressive strain (\( a = 3.38~\text{\AA} \)), enhanced Fe–chalcogen hybridization amplifies the local exchange field, resulting in increased \( \Delta_s \). 
In contrast, tensile deformation (\( a = 4.13~\text{\AA} \)) reduces orbital overlap, producing a more isotropic spin distribution and suppression of the spin splitting. 
Despite these structural distortions, the system retains its metallic character across the entire strain range.

Since spin--orbit coupling is not included in these calculations, the observed spin texture and finite \( \Delta_s \) originate solely from exchange interactions modulated by the Janus chemical asymmetry between Se and Te (and S) layers. 
The strain-dependent evolution of the spin density thus provides compelling microscopic evidence of strain-tunable altermagnetism in Janus FeSeTe (and FeSeS), confirming that the spin splitting arises from the exchange field associated with the anisotropic spin texture rather than from relativistic effects. 
Comparable strain-driven modulations of spin density and exchange anisotropy have been reported in other altermagnetic materials subjected to external perturbations~\cite{Strain-induced,Strain-induce-am,Strain-engineering,Strain-modulated,Strain-dp-am}.

When SOC is included in the calculations, both Janus FeSeTe and FeSeS monolayers lose their metallic character and evolve into insulating states, as shown in Figures S3 and S4 (see  the Supplemental Material \cite{SuppMat}). The SOC interaction opens a finite gap at the Fermi level, with a smaller induced gap observed in FeSeS due to its weaker relativistic effects. 
Nevertheless, the systems retain their exchange-driven altermagnetic character, as the dominant spin splitting persists and remains governed by the anisotropic exchange field rather than SOC-induced interactions.

\begin{figure*}[!t]
 	\centering
 	\includegraphics[width=0.98\textwidth]{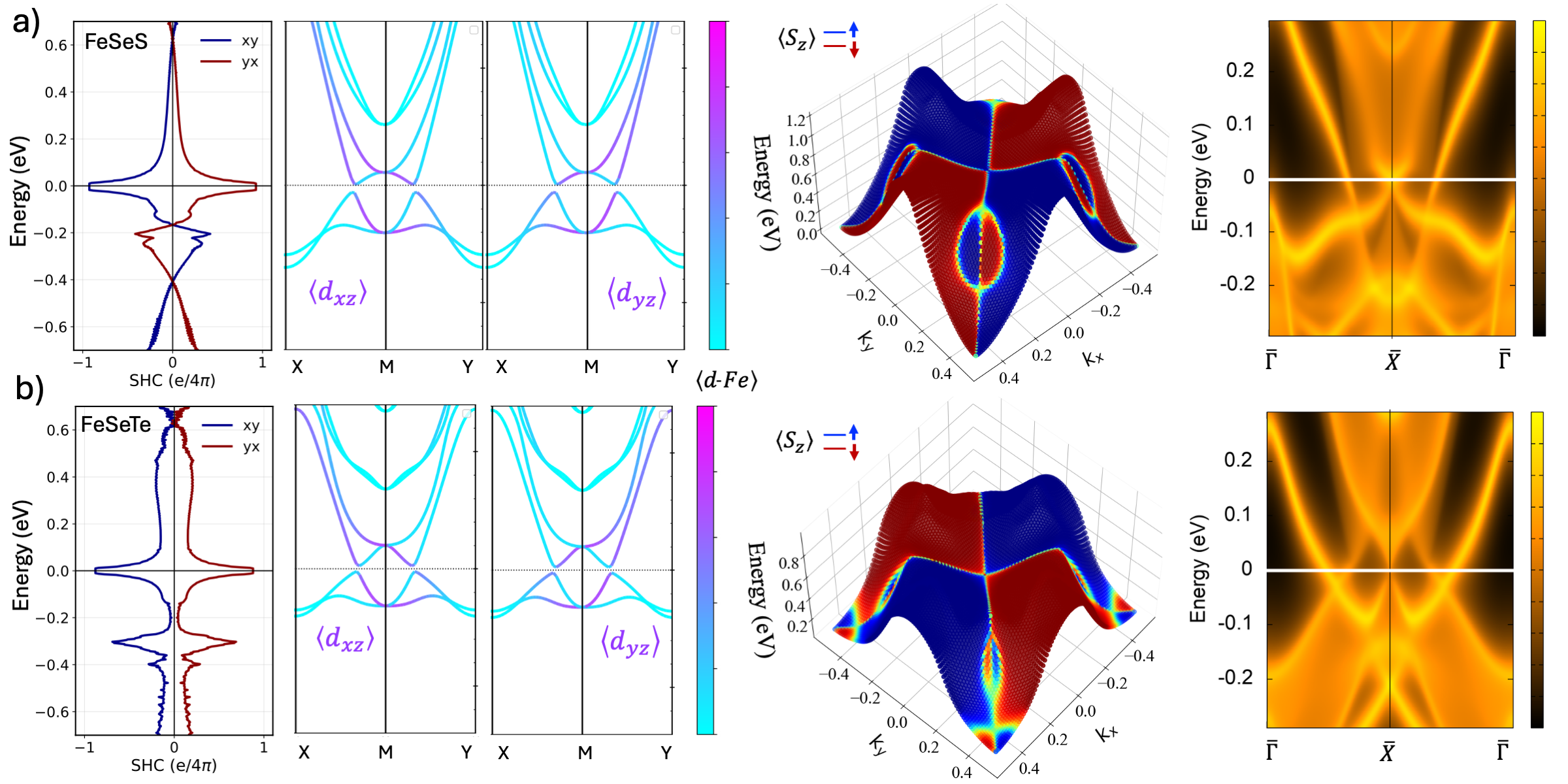}
 	\caption{(Color online) Spin–orbit–coupled electronic and topological properties of FeSeX (X = S, Te) Janus monolayers. (a)–(b) From left to right: intrinsic spin Hall conductivity (SHC) tensor components $\sigma_{xy}^{z}$ and $\sigma_{yx}^{z}$ as a function of the Fermi level; spin–orbit–coupled band structures projected onto Fe $d_{xz}$ and $d_{yz}$ orbitals; three-dimensional first conduction band dispersion colored by the spin expectation value $\langle S_z \rangle$; and the corresponding surface (010) band structure along the $\bar{\Gamma}$–$\bar{X}$–$\bar{\Gamma}$ path. The orbital-resolved bands highlight the dominant contribution of Fe $d_{xz/yz}$ states to the low-energy manifold, which are responsible for the SOC-induced topological gap near the Fermi level. The opposite signs of $\sigma_{xy}^{z}$ and $\sigma_{yx}^{z}$ reflect the preserved $C_{4z}T$ symmetry, while the surface states reveal gapless edge modes within the two-dimensional energy gap, confirming the quantum spin Hall insulating nature of FeSeS and FeSeTe.} 
 	\label{fig:SHC}
 \end{figure*}

Comparison of band structures with and without SOC for Janus FeSeS and FeSeTe (Figure S5 \cite{SuppMat}) shows that the momentum-dependent altermagnetic spin splitting \( \Delta_s \) near the M-point, including its $d$-wave pattern, is already present without SOC and remains essentially unchanged when SOC is included, confirming its exchange-driven origin, while SOC acts primarily to gap band crossings and induce band topology on top of the altermagnetic electronic structure. In addition, DFT+U+SOC calculations for Janus FeSeS and FeSeTe demonstrate that the SOC-induced energy gap persists over a relevant range of on-site Coulomb interactions (Figure S7 \cite{SuppMat}).

\subsection{Topological properties}

The spin–orbit–coupled electronic structure of the Janus FeSeX (X = S, Te) monolayers at their equilibrium lattice constants is depicted in Figure~\ref{fig:SHC}. The inclusion of spin–orbit coupling opens a direct energy gap of approximately 34 $m$eV in FeSeS and 24 $m$eV in FeSeTe, with both the valence-band maximum and conduction-band minimum located near the M point along the X–M–Y high-symmetry lines. The electronic states around the Fermi level are dominated by $d$-Fe orbitals, particularly the ($d_{xz}$) and ($d_{yz}$) components, which interchange their orbital character along the X–M and Y–M directions, respectively. This orbital interchange reflects the underlying ($C_{4z}T$) symmetry, giving also rise to an orbital altermagnetic texture in momentum space, as has been predicted in tight-binding models \cite{leeb2024spontaneous,orbital-spin-AM}.

The spin-resolved conduction-band structure reveals an opposite spin splitting along the $\Gamma$–X and $\Gamma$–Y directions, which persists even in the presence of spin–orbit coupling, while band degeneracy is preserved along $\Gamma$–M, consistent with the constraints imposed by the $d$-wave altermagnetic symmetry operation $C_{4z}T$.

Upon the inclusion of SOC, a band inversion occurs between the Fe ($d_{xz}$) and ($d_{yz}$) orbitals through the energy gap induced by SOC, indicating the emergence of a non-trivial topological phase (see Figure ~\ref{fig:SHC}). Hence, the SOC-induced gap in both FeSeS and FeSeTe exhibits a topological character, where the inversion of the orbital band produces an intrinsic topological insulating state protected by the magnetic symmetry $C_{4z}T$.

The appearance of a topological energy gap is evidenced by the quantized behavior of the spin Hall conductivity (SHC) within the insulating region (see left panels of Figure ~\ref{fig:SHC}). Within this energy window, the SHC takes a constant value near $e/4\pi$. The observed opposite signs of $\sigma_{xy}^z$ and $\sigma_{yx}^z$ are indicative of the response enforced by $C_{4z}T$ symmetry.  
The topological character of this insulating phase is further confirmed by the calculated spin Chern number ($\mathrm{SCN}=1$) and $\mathbb{Z}_2$ index ($=1$), both confirming the presence of a nontrivial topological order. The quantized SHC plateaus observed within the SOC-induced gap provide evidence of a two-dimensional quantum spin Hall state, protected by the $C_{4z}T$ symmetry \cite{modelhamiltonianAM,serrano2025}.  

The surface energy bands along the (010) direction (right panels) display gapless edge modes bridging the valence and conduction bands, 
confirming the presence of topologically protected edge states that give rise to robust conduction channels at the sample boundaries. These results establish Janus FeSeX (X = S, Te) monolayers as a unique class of $d$-wave altermagnetic topological insulators, where spin–orbit coupling and altermagnetic symmetry intertwine to produce a Hall state of quantum spin protected by $C_{4z}T$  symmetry.

\begin{figure}[h]
	\center
	\includegraphics[width=0.50\textwidth]{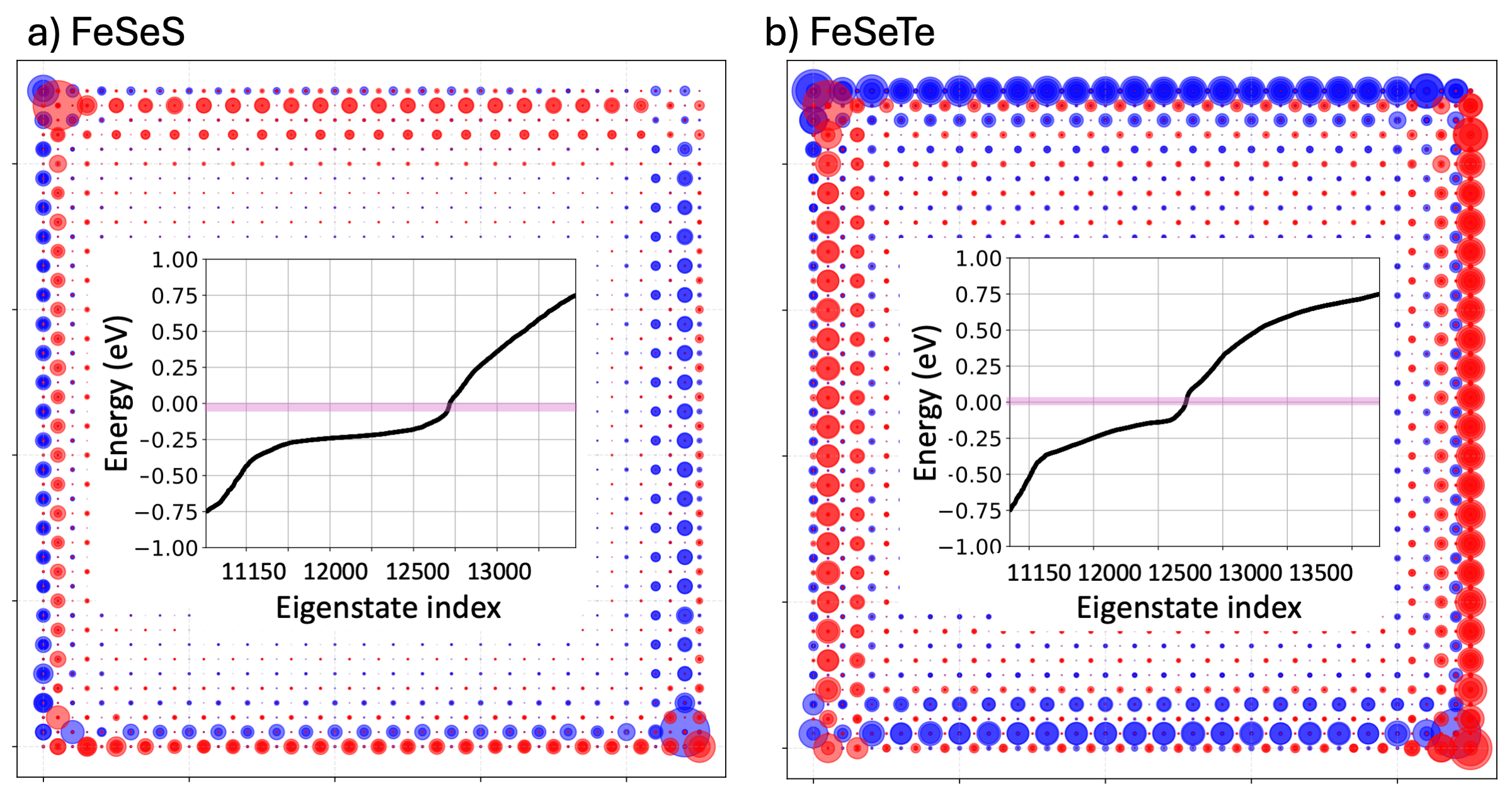}
	\caption{Spin-polarized edge modes in FeSeX (X = S, Te) Janus square geometries obtained from finite-size calculations. The red and blue colors represent spin-up and spin-down wave function densities localized along the system boundaries. The inset presents the energy spectrum of a 23×23 finite square lattice versus eigenstate index, with edge states emerging within the 2D energy gap (shaded region).}
	\label{fig:SPE}
\end{figure}

To further establish the edge-localized nature of the topological states in Janus FeSeX (X = S, Te) monolayers, we analyzed the real-space wave function amplitudes for the in-gap modes obtained from the finite-size calculations. 
These states are found to be localized at the boundaries of the 2D  system providing direct real-space evidence of topological edge states. 
Figure~\ref{fig:SPE} shows the spin-resolved edge states, with spin-up (red) and spin-down (blue) channels localized at the boundaries of the square geometry. 
The corresponding finite-size energy spectrum of the finite system, plotted as a function of eigenstate index, exhibits energy states within the 2D band gap. 
The combined energy spectral and real-space wave function distribution confirms the boundary origin of these modes, consistent with the quantum spin Hall phase.
These findings verify the coexistence of altermagnetic order and quantum spin Hall topology in Janus FeSeX (X = S, Te) monolayers.

\section{CONCLUSIONS}

This study reveals that Janus FeSeX  monolayers (X = S, Te) represent a compelling paradigm for realizing intrinsic topological altermagnetism, where the interplay between symmetry breaking, magnetic interactions, and relativistic effects yields a quantum material. 
The chemical asymmetry inherent to the Janus configuration systematically breaks out-of-plane mirror and inversion symmetries, reducing the crystalline point group to \textit{P}4mm, and thereby enabling momentum-dependent spin splitting, characteristic of altermagnetic phases. 
Our first-principles calculations demonstrate that these symmetry modifications, combined with the exchange interactions, give rise to a compensated magnetic state, exhibiting anisotropic spin textures, a hallmark of altermagnetism.

The sensitivity of  exchange splitting $\Delta_{s}$ to biaxial strain underscores the tunability of the electronic and magnetic properties within these systems, providing a promising avenue for strain-engineered spintronics applications. 
Specifically, strain enhances or suppresses the exchange-driven spin splitting without compromising the dynamical stability of the stable monolayers, as confirmed by phonon dispersion analyses. This strain-tunable altermagnetic behavior aligns with previous theoretical proposals for symmetry-controlled spin textures \cite{Gomonay2024}.

Beyond magnetic order, the inclusion of spin--orbit coupling (SOC) induces a finite topological gap ($\sim 24$ to $34\, \mathrm{meV}$), stabilizing quantum spin Hall  states characterized by non-trivial $\mathbb{Z}_{2}$ invariants and quantized spin Hall conductivity. 
Importantly, the coexistence of $d$-wave altermagnetic order and non-trivial topology highlights a novel phase of matter, \textit{topological altermagnets}, where magnetic order does not suppress but coexists with topologically protected edge modes.
These edge modes exhibit robustness against non-magnetic disorder, suggesting potential for dissipationless spin transport at ambient conditions.

The sizable SOC-induced gaps are particularly promising for practical applications in spintronics and quantum information, where topological protection can enable low-dissipation devices operational at or near room temperature. 
Moreover, the symmetry-protected nature of the spin textures and topological states suggests that external perturbations such as electric fields or substrate interactions could further modulate these phases, offering additional degrees of control.

\vspace{1cm}

\section*{Acknowledgments}

R.G.-H. gratefully acknowledges continuous support from the Alexander von Humboldt Foundation. The authors would like to thank the HCP-Granado Uninorte Cluster for providing computational resources.

\bibliographystyle{apsrev4-2-titles}
\bibliography{library}

\end{document}